\documentclass[aps,prd%,showkeys
,showpacs,amssymb,
amsfonts,epsf,preprintnumbers,nofootinbib,superscriptaddress]{revtex4}
\usepackage[dvips]{graphicx}
\usepackage{bm,latexsym,amsmath,amssymb,amsfonts,color}
\usepackage{enumerate}
\begin{document}

\title{Causality Problem in a Holographic Dark Energy Model}

\author{Hyeong-Chan Kim}
\email{hckim@ut.ac.kr}
\affiliation{School of Liberal Arts and Sciences, Korea National University of Transportation,
Chungju, 380-702, Korea}
\author{Jae-Weon Lee}
\email{scikid@jwu.ac.kr}
\affiliation{Department of energy resources development, Jungwon University, 5 dongburi, Goesan-eup, Goesan-gun Chungbuk, 367-805, Korea}
\author{Jungjai Lee}
\email{jjlee@daejin.ac.kr}
\affiliation{Department of Physics, Daejin University, Pocheon, 487-711, Korea}

\begin{abstract}
In the model of holographic dark energy,
there is a notorious problem of circular reasoning between the introduction of future event horizon and the accelerating expansion of the universe.
We examine the problem after dividing into two parts, the causality problem of the equation of motion and the circular logic on the use of the future event horizon.
We specify and isolate the root of the problem from causal equation of motion as a boundary condition, which can be determined from the initial data of the universe.
We show that there is no violation of causality if it is defined appropriately and the circular logic problem can be reduced to an initial value problem.
\end{abstract}

%\preprint{CQUeST 2009-0XXX}
\maketitle

To explain the present accelerating expansion of the universe discovered in 1998~\cite{dE}, several dark energy models have been suggested by introducing new exotic matters or by modifying gravity.
For the reviews about the accelerating expansion and dark energy, consult 
Ref.~\cite{dEmodels}.
Of all the models, the holographic dark energy (HDE) model takes a unique position because it is an effective theory which introduces an energy density determined by geometric structures of the universe. % of which physical origin is unknown. 
Interestingly, to explain the present accelerating expansion of the universe, it was shown that the theory requires the future event horizon as the boundary of universe rather than the Hubble, the particle, or the apparent horizons~\cite{Li}.
There are several attempts to search for the origin for the HDE from Casimir energy in de Sitter space~\cite{casimir}, quantum uncertainty of transverse position~\cite{age1}, holographic gas~\cite{hologas}, entanglement entropy from quantum information loss~\cite{jkj}, the Hawking radiation from the future event horizon~\cite{HawkingRad}, or the space-time curvature~\cite{curvDE}.
However, many physicists doubt that most of them may violate the causality or have a circular reasoning problem since the future event horizon is defined globally~\cite{age1}.

The problems go as follows:
\begin{itemize}
\item {\it Circular logic}:
%The future accelerating phase is essential to the existence of future event horizon, whose existence is indispensable to the HDE.
%Therefore it is said that the future accelerating expansion is inevitable in the presence of the HDE.
%
Given the present data of the universe, we do not know whether the universe will eventually undergo accelerating expansion or not.
 The future event horizon is determined only after the evolution of the universe finished. %, one can determine whether the event horizon exists or not.
  On the other hand, once the HDE is introduced the universe is destined to expand with accelerating rate.
%  If we do not introduce the HDE, the universe evolves through the Einstein equation including some mysterious energy source and may or may not have a future event horizon.
%  
%Therefore, introducing the HDE appears to determine the future of the universe by hand.
 Then, can we use the future event horizon even if we do not know its existence at present?
  This composes the heart of the circular logic.
  The problem becomes worse if we note that the holographic energy based on other horizons than the future event horizon does not accelerate the expansion of the universe.

\item {\it Causality problem}:  Assume that there is a creature which can modify the future event horizon.\footnote{Here we disregard the fact that it is impossible to modify the future event horizon at present because it is determined by the whole evolution of the universe including the creature's action.}
    If the creature modifies the horizon, it may affect the present motion of the universe since the equation of motion depends on the distance to the future event horizon.
    This raises the problem on causality: ``Can we expect the next second state of the universe from the present data in the absence of future knowledge?"

\end{itemize}
To overcome the problem of circular reasoning, Gong~\cite{gong} developed the extended holographic dark energy model, where the Brans-Dicke theory of gravity is adopted using the Hubble scale in place of the future event horizon.
They have shown that there exists a no-go theorem that the Hubble scale cannot be an infrared cutoff for the universe with Brans-Dicke gravity.
However, in the presence of a potential term for the Brans-Dicke scalar field, they succeed to show that it is possible to generate the HDE from the Hubble horizon~\cite{Liu}.

In the present work, we directly analyze the HDE rather than introducing another theory of gravity or matters.
%
%According to the standard cosmological model, most of the universe is composed of the cosmological constant and cold dark matters.
%The cosmological constant forces the space-time to have a 
In the spatially homogeneous and isotropic universes, the future event horizon is located at a distance
\begin{eqnarray}
R_h =a(\tau)\int_\tau^\infty \frac{d\tau'}{a(\tau')}, \label{EH}
\end{eqnarray}
where $\tau$ is the comoving time and $a(\tau)$ is the scale factor of the Robertson-Walker metric
\begin{eqnarray}
ds^2 = -d\tau^2+a^2(\tau)d{\rm \bf x}^2 \label{RW}\,.
\end{eqnarray}
Note that the future horizon satisfies $\dot R_h/H = R_h -H^{-1}$.
Therefore, the horizon is always located outside the Hubble (apparent) horizon if $\dot R_h> 0$.
Any observers in the universe will be surrounded by the future event horizon and their accesses to remote information of the universe are restricted by the horizon.
This lack of information is represented by a kind of entropy given by the cosmological horizon area in Planck unit similar to the black hole entropy.
The generalized second law of thermodynamics for the Robertson-Walker space-time was proved by Daives~\cite{Davies} and Pollock and Singh~\cite{Pollock} showing that the total entropy of gravity+matter does not decrease through physical processes.

How can we predict the future of the universe?
At present, we are observing the universe inside the past light-cone of us.
At every moments, our range of observation increases with time and new data on the universe arrive us.
Therefore, in principle, we cannot predict anything for the future.
To circumvent this situation, we resorts to the observational data for the past.
From the observations, it was found that the observable universe is spatially homogeneous and isotropic in large scales.
%Based on this observations, the homogeneity and the isotropy of space are conjectured, which are two principles of physical cosmology.
%Even though human cannot observe outside the boundary of our past, one expects that the outside will not be much different from the inside in large scales.
%In this sense, 
We assume in this work that our universe is spatially {\it homogeneous} in large scale.

Many dark energy models were appeared in the literatures~\cite{remodels} to explain the recent accelerating expansion of the universe.
Some of them introduces exotic matters such as Chaplygin gas, phantom matter, quintessence, or others.
Some others modify the gravity theory by considering higher curvatures or branes in higher dimensions.
The holographic dark energy locates in the middle of the two directions because it introduces an energy density determined by a geometric quantity.
There are several versions of holographic dark energy models.
Li first introduced the future event horizon to give the future acceleration~\cite{Li}.
Some authors explain the origin of the holographic dark energy as the quantum energy fluctuation~\cite{pad06}, cosmic Hawking radiation~\cite{kll08}, e.t.c.
Many variations of the holographic dark energy models are also studied including agegraphic dark energy model~\cite{age1,agegraphic} and Ricci dark energy models~\cite{curvDE}.

The equation of motion in the presence of the HDE is given by,
\begin{eqnarray}\label{eom}
3 M_p^2H^2 = \sum \rho =\rho_h+\rho_{nh} ,
\end{eqnarray}
where $H$ denotes the Hubble parameter. 
The energy densities are divided into two pieces the holographic dark energy, $\rho_h$ and the sum of all energy densities other other than the HDE, $\rho_{nh}$.
The portion of the HDE in our universe becomes
$$
\Omega_h = \frac{\rho_h}{\sum \rho } = \frac{\rho_h}{3M_p^2 H^2}= \frac{d^2}{(HR_h)^2}\,,
$$
where $d$ is a constant depending on the present HDE density.

The causality problem of HDE was recognized by Li~\cite{Li} for the first time when he introduces the future event horizon.
He also provided part of the solution by questioning about the usefulness of co-moving time.
Even though the co-moving time is intrinsic to a comoving observer in a time-dependent background, it may not be the best time-parameter to use in order to understand the causality.
He show that the event horizon is no-longer as acausal as in the co-moving time if we write the metric~(\ref{RW}) in the conformal time, $\eta = \int_\infty^\tau \frac{d\tau'}{a(\tau')}$, as
$$
ds^2 = a^2(\eta)(-d\eta^2 + dr^2 + r^2d\Omega^2_{2}).
$$
The range of the conformal time has a finite upper limit, for instance $\eta \in (-\infty, 0)$.
Due to this finite upper limit, a light-ray starts from the origin at the time $\eta$ can not
reach arbitrarily far but have a horizon at $r = -\eta$.
Now, the formula for the horizon distance $R_h = a(\eta)|\eta|$ appears to be causal.
He also mentioned about the possibility that how the quantized vacuum energy in a box can be interpreted as the holographic dark energy.

Li's resolution present a way to avoid the causality problem.
In this work, we extend his resolution to general situations without introducing {\it ad hoc}
conformal time, which is the definite integral of the inverse of scale factor from infinity to $\tau$. 
In addition, we discuss the conceptual aspect of the future event horizon.

\subsubsection{Does the future event horizon always leads to future acceleration?}
First, we provide a counter example to the assertion: ``If there exists an % matter field with its
energy density depending on the future event horizon, then the universe is destined to expand with accelerating rates in the future." 
%We provide a counter example to the assertion.
Let us assume an energy density $\rho_h=3 M_p^2 p^4 R_h^n$ with an integer $n$ and a real number $p$.
If $n=-2$, it is nothing but the holographic dark energy, which is known to give future accelerating expansion respecting the assertion.
At the present case, let us consider the case with $n=2$.
In the absence of other field (or if the field dominates the universe which will happen for large $R_h$)
the equation of motion becomes $H^2 = p^4 R_h^2$.
Since we are interested in finding an expanding solution we may reduce the equation
$$
\frac{\dot a}{a^2} =  p^2 \int_\tau^\infty \frac{d\tau'}{a(\tau')},
$$
and then differentiating both sides with respect to $\tau$ once, we get
%\begin{eqnarray*}
$\left(\frac{d}{d\tau}\right)^2 \frac{1}{a} = \frac{p^2}{a}$.
%\end{eqnarray*}
Note that its general solution
$$
 a(\tau) = (\alpha e^{p \tau}+\beta e^{-p \tau})^{-1},
$$
where $\alpha$ and $\beta$ are arbitrary integration constants,
does not lead to future accelerating universe contrary to the assertion.
In this case, the initially expanding universe eventually contracts and there is no accelerating expansion in the future.
Therefore, the energy density depending on the future event horizon does not necessarily give the future accelerating expansion of the universe.

\subsubsection{Causal part and acausal part of the evolution equation }
Now we specify the root of the causality problem in the evolution equation~(\ref{eom}).
Then, we isolate it from well posed differential equation as a boundary condition, which can be determined from initial data.
It appears that the equation bears the causality problem in the following sense:
Suppose that there exists a creature which can modify the future event horizon.
Someday in the future, the creature decides to change the horizon.
Now, we should ask the following questions:
 \begin{enumerate}
 \item Does this action modify the present evolution of the universe?
 \item Does this imply the causality violation for us?
 \end{enumerate}
For the first question, we should say ``yes."
Noting the evolution equation~(\ref{eom}), it is quite sure that the modification of the horizon area actually changes the present evolution of the universe.
However, this does not mean the answer to the second question is also yes.

To answer the second question, we examine the equation~(\ref{eom}) in detail.
In fact, we separate the future dependent part from the real dynamics which is described by a well posed second order differential equation.
To do this, we rewrite the equation of motion as
$$
  \int_{\tau}^\infty \frac{d\tau'}{a(\tau')}
   = \frac{d}{a(\tau) \sqrt{H^2 -\frac{\rho_{\rm nh}}{3 M_p^2}}},
$$
where $\rho_{nh}$ implies the sum of all energy densities other than the holographic dark energy and we assume $d>0$.
Note that the future dependent part is localized in the left hand side of the equation.
We can separate that part from others by setting $\int_{\tau}^\infty \frac{d\tau'}{a(\tau')}= r_\infty -\int_0^\tau\frac{d\tau'}{a(\tau')}$, where  $r_\infty=\int_0^\infty \frac{d\tau}{a(\tau)}$ bears the future dependence of the equation of motion.
If we differentiate above equation once with respect to $\tau$, the term $r_\infty$ disappears and obtain a well posed second order differential equation:
\begin{eqnarray} \label{eom:2}
\dot H - H^2 +\frac{1}{3 M_p^2} \left(\rho_{\rm nh} -\frac{\dot \rho_{\rm nh}}{2 H}\right) -\frac{(H^2-\rho_{\rm nh}/(3M_p^2))^{3/2}}{d H}=0.
%
%    -\frac{1}{a(\tau)} = \frac{d}{d\tau} \frac{\mathfrak{F}(\dot a(\tau), a(\tau))}{a(\tau)}.
\end{eqnarray}
In the series of calculations, we divide the evolution equation~(\ref{eom}) into two pieces, one is a well posed evolution equation~\eqref{eom:2} and the other is the term which appears to bear the information of the causality violation,~$r_\infty$.

Note that the value $r_\infty$ does not affect on the evolution equation~(\ref{eom:2}).
The whole evolution can be solved from the information at some initial time without any future information.
Since the evolution can be determined in the absence of $r_\infty$, we find that the horizon can be determined from the information at present if we solve the equation of motion~(\ref{eom:2}).
In this sense, the information about $r_\infty$ needs not be specified to determine the distance to the event horizon.
In fact, $r_\infty$ is not always well defined.
It becomes a finite number only when the scale factor behaves as $a(\tau)\sim \tau^m$ $(m < 1)$ for small $a$ and $a(\tau) \sim \tau^n$ $(n>1)$ in the future.
However, even if $r_\infty$ is ill-defined, the evolution equation~(\ref{eom:2}) works well reproducing the Einstein equation.
If someone says that the causality is violated, he means that something unexpected from the past data happens at present (due to future action).
However, as seen in Eq.~(\ref{eom:2}), all future evolutions can be predicted from the past data.
If we define the `causality' in the sense that there does not appears which is forbidden from the past data, then we may say that the answer to the second question is ``No", there is no `causality' violation for us.

It is interesting to see the role of $r_\infty$.
Even though $r_\infty$ is defined by the integral $\int_0^\infty a^{-1}d\tau$, we may also obtain its value by using the limit
\begin{eqnarray} \label{r:infty}
r_\infty =\lim_{\epsilon \to 0} \int_{\epsilon}^\infty \frac{d\tau'}{a(\tau')}=\lim_{\epsilon \to 0} \frac{d}{\sqrt{\dot a^2- \frac{a^2\rho_{\rm nh}}{3M_p^2} }}.
\end{eqnarray}
Therefore, one may simply take the limit to obtain $r_\infty$ rather than integrating over the whole evolution.
From this point of view, the value $r_\infty$ does the role of initial {\it boundary condition} rather than a future dependent constant.
Summarizing, if the creature succeeds to modify the future event horizon, it simply means that it manages to modify the boundary condition or the initial condition given by $r_\infty$.
Since the universe follows the evolution equation~(\ref{eom:2}) with the modified initial condition from the beginning, any observer including the creature may not notice the change of the initial condition since he or she has been living in the universe with modified initial condition.
In this sense, there happens no causality violation from the point of view of an observer in a universe.

\subsubsection{Resolution of the circular logic}
The origin of the circular logic problem comes from the ignorance of the existence of the future event horizon.
However, note that the cosmological event horizon cannot be created or be removed by any classical means.
The existence or the absence of the horizon is determined from the beginning of the universe.
Thus, the cosmological solution can be divided into two separated classes: the universe with future event horizon and the universe without future event horizon.

Can we find out where we are living of the two classes?
As shown in the previous calculations, the evolution equation~(\ref{eom}) can be modified into one second order differential equation~(\ref{eom:2}) supplemented by a boundary condition~(\ref{r:infty}) which can be determined at the beginning of the universe.
Given the present data of the universe, we may trace back the universe to know the initial situation by using Eq.~\eqref{eom:2}.
By doing this procedure, the boundary condition $r_\infty$ can be obtained from the limiting procedure~(\ref{r:infty}).
Note that, in doing this procedure, we do not have to know any future information.

Even when we do not know $r_\infty$, we can predict the future from the evolution equation~(\ref{eom:2}) since it is a well posed differential equation.
Therefore, we may notice where we live of two classes in principle, if we know the present data of the universe in detail.
The above logic holds only when the assumption of space homogeneity is valid in our universe since we predict the future evolution on the basis.
Still, we cannot differentiate whether we are living in a universe with future event horizon or without until we have more precise data of the present universe including dark matters and geometries.
However, the use of Eq.~(\ref{eom}) for the next moment evolution can be justified since all of the future evolution is governed by the well posed second order differential equation~(\ref{eom:2}) and the future evolution is only related with the boundary condition $r_\infty$ which can be determined at the beginning of the universe.

\subsubsection{Discussions}

We studied the problem of circular reasoning between the introduction of future event horizon and the accelerating expansion of the universe in HDE. 
We divided the problem into the causality of the equation of motion and the circular logic on the use of the future event horizon. 
The root of the causality problem was identified as a number $r_\infty$ and was separated from  a causal equation of motion~\eqref{eom:2}.
The explicit value of $r_\infty$ can be determined at the beginning of the universe. 
We have shown that the causality problem is absent if we require the causality in a weak sense that any event prohibited from the present data will not happen in the future. 
We also reduced the circular logic problem to the initial condition, $r_\infty$, by noting the fact that the existence (or absence) of the cosmological event horizon is determined from the initial data of the universe, which can be traced back from the precise present data.

The philosophy of Einstein equation is that ``matter determines the geometry and the geometry rules the motion of the matter".
At the present case, $r_\infty$ has a geometric origin since it is determined from the motion of the scale factor.
Therefore, the holographic dark energy is originated from geometry and also affects on the motion of the scale factor.
As mentioned by Li~\cite{Li}, it still appears rather puzzling why holographic energy is given by the time-dependent horizon size related to this $r_\infty$.
In any cases, the holographic dark energy model must be an effective description based on fundamental principles such as UV/IR connection~\cite{Cohen}, bulk holography~\cite{Thomas}, space-time foam~\cite{Ng,Myung}, quantum fluctuation of the space-time itself~\cite{Gao} or others.

\begin{acknowledgements}
HCK was supported in part by the Korea Science and Engineering Foundation
(KOSEF) grant funded by the Korea government (MEST) (No.2010-0011308) and the APCTP Topical Research Program (2012-T-01).
HCK personally thanks to the CQUeST for its warm hospitality during my visiting.
\end{acknowledgements}

%\appendix

\end{document}